\documentclass{article}
\usepackage{spconf,amsmath,graphicx,hyperref}
\usepackage{booktabs}
\usepackage{threeparttable}

\title{A Study of Data Selection Strategies for Pre-training Self-Supervised Speech Models}

\name{Ryan Whetten$^1$, Titouan Parcollet$^2$, Marco Dinarelli$^3$, Yannick Estève$^1$\thanks{This work received funding from the French ANR E-SSL project (N°ANR-22-CE23-0013) and used HPC resources from GENCI–IDRIS (AD011014732 and A0131013821)}}
\address{$^1$Laboratoire Informatique d'Avignon, Avignon Université, Avignon, France \\
$^2$University of Cambridge, Cambridge, United Kingdom \\ 
$^3$Laboratoire d'Informatique de Grenoble, Université Grenoble Alpes, Grenoble, France
}
\begin{document}
%
\maketitle
\begin{abstract}
Self-supervised learning (SSL) has transformed speech processing, yet its reliance on massive pre-training datasets remains a bottleneck.
While robustness is often attributed to scale and diversity, the role of the data distribution is less understood.
We systematically examine how curated subsets of pre-training data influence Automatic Speech Recognition (ASR) performance.
Surprisingly, optimizing for acoustic, speaker, or linguistic diversity yields no clear improvements over random sampling.
Instead, we find that prioritizing the longest utterances achieves superior ASR results while using only half the original dataset, reducing pre-training time by 24\% on a large corpora. 
These findings suggest that for pre-training speech SSL models, data length is a more critical factor than either data diversity or overall data quantity for performance and efficiency, offering a new perspective for data selection strategies in SSL speech processing.
\end{abstract}
\begin{keywords}
self-supervised learning, speech, automatic speech recognition, data selection
\end{keywords}
\section{Introduction}
\label{sec:intro}
Self-supervised learning (SSL) is a framework for training neural networks where pseudo-targets are generated from the input data itself. A neural network model is trained to using these targets in a stage called pre-training, and then can be subsequently fine-tuned using human labeled data for a particular downstream task.
SSL in speech processing has led to state of the art performance in various tasks such as automatic speech recognition (ASR), taking advantage of massive amounts of unlabeled data to train a neural network to produce high quality speech representations~\cite{mohamed2022self}. 

To take advantage of this data, SSL speech models are large, typically composed of 12 or more Transformer based layers~\cite{baevski2020wav2vec}. While these frameworks have led to high performing models, they require a significant amount of resources in terms of the amount of memory, time, and data to train these large neural networks. 

To improve the efficiency of SSL speech models, researchers have focused on improving model architectures~\cite{whetten2024analysis, baevski2023efficient}, simplifying or adjusting pre-training objectives~\cite{chen2023reducing, pmlr-v162-chiu22a, baevski2023efficient}, and proposing efficient evaluation methods~\cite{whetten25_interspeech}. 
While these methods reduce memory requirements and the time it takes to train and evaluate, high performing SSL speech models such as Google USM~\cite{zhang2023google} and Xeus~\cite{chen-etal-2024-towards-robust} are still pre-trained on the order of millions of hours of speech data (corresponding centuries of audio), and the effects of changes in the composition of pre-training data has been less explored.

To address this gap, we investigate the impact of simple unsupervised data selection strategies for pre-training. Our first set of methods samples data to promote diversity across acoustic, speaker, or linguistic features, while our second set focuses on sampling based on utterance length.

Unexpectedly, our results show that enforcing diversity in the pre-training data—whether acoustic, speaker, or linguistic—yields no significant improvement in downstream Automatic Speech Recognition (ASR) performance. In contrast, simply selecting the longest utterances for pre-training achieves superior ASR performance while requiring only half of the original dataset and reducing pre-training time by 24\% on a large-scale corpus. 
We make code used in the experiments public for reproducibility\footnote{link to code will be provided for camera ready version}.

\section{Related Work}
\label{sec:related-work}
In this section, we provide an overview of prior work on improving the efficiency of SSL for speech models, as well as studies on data selection for speech processing.

\subsection{Efficiency for SSL speech models}
\label{sec:efficiency-ssl}
Research on addressing the inefficiency of SSL speech models has primarily focused on three directions: modifications to the SSL training framework, architectural innovations, and alternative evaluation methodologies.

For improving the SSL framework itself, researchers have explored simplifying the generation of pseudo-targets and objectives. For example, BEST-RQ~\cite{pmlr-v162-chiu22a} employs a random-projection quantizer and a simple cross-entropy loss, in contrast to wav2vec~2.0~\cite{baevski2020wav2vec}, which uses a learned codebook along with a combination of contrastive and diversity losses. Other modifications involve leveraging pre-trained or ASR fine-tuned models to generate higher-quality embeddings, which serve as training targets and yield both performance and efficiency gains~\cite{chen2023reducing}. Additionally, practical engineering improvements to the SSL framework have been shown to significantly reduce memory usage and training time such computing targets on the fly to reduce storage memory as and not encoding the masked tokens~\cite{chen2023reducing, baevski2023efficient}.

Architectural changes have also played a key role in improving efficiency. In models such as BEST-RQ~\cite{pmlr-v162-chiu22a} and Whisper~\cite{radford2023robust}, the acoustic feature extractor has been simplified, using mel-filterbanks with only two convolutional layers instead of raw audio inputs with seven convolutional layers. Other work has investigated replacing multi-head self-attention blocks with more efficient alternatives~\cite{whetten2024analysis}. In parallel, knowledge distillation methods have been explored to reduce model size while maintaining strong performance~\cite{chi25_interspeech}.

Finally, researchers have developed more efficient evaluation approaches for SSL speech models. Recent work demonstrates that the rank of embeddings produced by an SSL model can serve as an indicator of pre-training quality~\cite{whetten25_interspeech, 10889651}. While these lines of research address efficiency from multiple perspectives, relatively little attention has been given to improving data efficiency in pre-training.



\subsection{Data selection for speech models}
\label{sec:related-work}

While substantial work has been done on data selection for speech models, most prior research has focused on active learning. Active learning involves identifying the most informative data for labeling, with the goal of minimizing annotation costs. Typically, this is achieved by using uncertainty or confidence measures, where utterances with the lowest confidence are selected for labeling and then used to train a fully supervised ASR model~\cite{malhotra19_interspeech}. In contrast, our work focuses on understanding what constitutes an effective pre-training dataset for SSL speech models, rather than on active learning.

For SSL-specific data selection, researchers have proposed methods that select pre-training data most similar to the target downstream domain~\cite{lu22_interspeech}. 
One such approach trains language models on quantized targets of speech from both the pre-training pool and the target domain dataset, and then computes a normalized difference in log-probability between the two models. 
This score serves as an indicator of how similar a pre-training utterance is to the target domain. While effective for domain-matching, this line of work is distinct from ours, as we instead investigate the general properties of pre-training data that lead to strong overall performance, independent of any specific target domain.

Most closely related to our study is the work of Berrebbi et al.~\cite{berrebbi2023more}, who analyze the effects of varying the number of speakers and the amount of audio in pre-training datasets. Their results highlight the importance of dataset size for SSL, though their experiments were limited to under one thousand hours of pre-training data and require speaker labels. By contrast, our study explores methods that do not require speaker labels, consider features beyond speaker information, and scale up to 25 thousand hours of audio.


\section{Methods}
\label{sec:methods}
In this section, we describe the dataset used, the different methods for data selection, and the training environment.

\subsection{Data}

For our experiments, we use the Loquacious dataset~\cite{parcollet25_interspeech}, which combines commercially usable English speech corpora. We select this dataset because it is both large-scale and challenging, containing 25,000 hours of diverse speech, including read, spontaneous, conversational, clean, and noisy recordings.

This diversity extends across all splits (train, development, and test), which is crucial for our study. Unlike prior work that focuses on selecting pre-training data to closely match the fine-tuning domain~\cite{lu22_interspeech}, our goal is to investigate which types of pre-training data yield the best overall performance across various speech conditions. For pre-training, we experiment with the medium split (2,500 hours) and the large split (25,000 hours). For fine-tuning Automatic Speech Recognition (ASR) models, we use the small split (250 hours).

\subsection{Data selection methods}
\label{sec:data-selction-methods}
We consider two baselines. The first, more challenging baseline uses the entirety of the data for pre-training, while the second is a random selection of 50\% of the data. We refer to these as the \emph{all} and \emph{random} baselines, respectively.

The next set of methods is based on diversity sampling across different feature types: acoustic, speaker, and linguistic. To capture basic acoustic characteristics of each utterance, we compute the first 13 Mel-Frequency Cepstral Coefficients (MFCCs) along with their first- and second-order derivatives. We then average across frames, producing a 39-dimensional vector per utterance. 
For speaker features, we extract 256-dimensional speaker embeddings using the WeSpeaker model~\cite{Wang2023} with pyannote~\cite{Bredin23}, which encodes speaker-specific characteristics from the audio.
For linguistic features, we use the SENSE model (Shared Embedding for N-lingual Speech and tExt)~\cite{mdhaffar2025sensemodelsopensource}, a high-performing, language-agnostic, open-source model that generates semantic embeddings from speech and text. This produces a 1024-dimensional vector representing the semantic content of each utterance.

To obtain diverse and representative samples, we apply stratified sampling guided by k-means clustering. After clustering the dataset using one of the feature representations, we sample a total of $N$ utterances with balanced coverage across all $k$ clusters. To avoid underrepresenting small clusters, we first ensure that each cluster contributes at least one data point. We then iteratively and evenly sample additional utterances from clusters until reaching the target size $N$.

This approach prevents larger clusters from dominating the sample and guarantees coverage of smaller ones. We set $N$ to approximately 50\% of the total utterances, adjusting slightly per method to keep the total audio duration comparable to the random baseline. For the Loquacious medium and large datasets, we use $k=150$ and $k=200$, respectively.

Finally, we explore two methods based on utterance length. The first simply selects the longest 50\% of utterances, denoted as \emph{length}. The second combines length and speaker diversity: we use the k-means clusters on speaker embeddings from the diversity sampling method, then instead of sampling randomly within clusters, we select the longest utterances from each cluster, denoted as \emph{speaker+len}. Due to limited resources, we only tested this feature combination, as preliminary results showed speaker diversity sampling was slightly more effective. We maintain a total audio duration approximately equal to the random baseline.

\subsection{Training settings}
For pre-training, we use the BEST-RQ framework and architecture, which employs a random-projection quantizer to generate pseudo-targets~\cite{pmlr-v162-chiu22a} and conformer layers as the model backbone. We adopt rotary positional embeddings~\cite{su2024roformer} and configure the model with 12 layers, a hidden size of 640, 8 attention heads, and a feed-forward network dimension of 2048, resulting in approximately 100M parameters. We select BEST-RQ due to its efficiency~\cite{whetten-2024-open}, strong performance~\cite{pmlr-v162-chiu22a,zhang2023google}, and its open-source implementation in SpeechBrain~\cite{whetten-2024-open,speechbrain}.
For fine-tuning for speech recognition, we adopt the fine-tuning SpeechBrain recipe for BEST-RQ which applys a feed-forward neural network with CTC loss. We set vocabulary size is set to 1,024  byte pair encoding tokens following~\cite{parcollet25_interspeech}.

We use NVIDIA A100 GPUs. To estimate total GPU time, we measure the duration required to complete 50,000 training steps on a system with 8 A100s, and then extrapolate to the full 200,000 steps. We apply dynamic batching, which groups utterances of similar lengths and adjusts batch sizes to maintain a consistent amount of audio per batch. The maximum duration per batch is capped at 800 seconds per GPU, yielding a total batch size of approximately 1.77 hours of audio across 8 GPUs.

We also employ dynamic chunking, which applies dynamic masks to both the attention mechanism~\cite{zhang2020unified} and the convolutional block~\cite{li2023dynamic}. This limits left and right context access in a randomized manner, effectively simulating both streaming and non-streaming conditions during training. While our study does not focus on streaming, prior work has shown that pre-training and fine-tuning with dynamic chunking improves performance for both streaming and non-streaming ASR~\cite{duret2025indomainsslpretrainingstreaming}. All reported results are obtained in a non-streaming setting.

We train all models for 200,000 steps with identical hyperparameters, varying only the pre-training subsets described in Section~\ref{sec:data-selction-methods}. To assess statistical reliability, we apply bootstrapping with 1,000 samples to estimate 95\% confidence intervals.

\section{Results}
\label{sec:results}

\begin{table}
\centering
\caption{Automatic Speech Recognition (ASR) results with different pre-training subsets from the Loquacious medium and large splits. All models are fine-tuned on the small split. The \emph{Time} and \emph{Data} columns report the amount GPU hours and pre-training audio used.}
\label{tab:results}
\vspace{0.5ex}
\begin{tabular}{lllcc}
\toprule
\textbf{Pre-train Split} & WER & WER & Time & Data \\
 Selection method & \emph{dev} & \emph{test} & (hrs)  & (hrs) \\
\midrule
\textbf{Medium Split} \\
All         & 18.48 & 19.39 & 195 & 2.50 k \\
Random      & 18.80 & 19.82 & 202 & 1.25 k \\
MFCC        & 19.05 & 19.98 & 202 & 1.24 k \\
Speaker     & 18.78 & 19.72 & 204 & 1.23 k \\
SENSE       & 19.23 & 20.39 & 199 & 1.24 k \\
Length      & \textbf{18.08} & 19.02$^{*\dagger}$ & 202 & 1.25 k \\
Speaker+Len & 18.16 & \textbf{18.97}$^{*\dagger}$ & 205 & 1.24 k \\
\midrule
\textbf{Large Split} \\
All         & 17.12 & 18.08 & 263 & 25.2 k \\
Random      & 17.53 & 18.54 & 214 & 12.6 k \\
MFCC        & 17.49 & 18.39 & 221 & 12.6 k \\
Speaker     & 17.26 & 17.97$^{*}$ & 214 & 12.6 k \\
SENSE       & 17.67 & 18.42 & 385 & 12.6 k \\
Length      & 16.76 & 17.77$^{*\dagger}$ & 200 & 12.6 k \\
Speaker+Len & \textbf{16.60} & \textbf{17.42}$^{*\dagger}$ & 201 & 12.5 k \\
\midrule
No pre-train~\cite{parcollet25_interspeech} & 22.30 & 23.80 & - & - \\
\bottomrule
\end{tabular}
\begin{tablenotes}
\footnotesize
\item $^*$Significantly lower WER from \emph{random} baseline ($p < 0.05$)
\item $^\dagger$Significantly lower WER from \emph{all} baseline ($p < 0.05$)
\end{tablenotes}
\end{table}

\begin{figure*}
\centerline{\includegraphics[width=\textwidth]{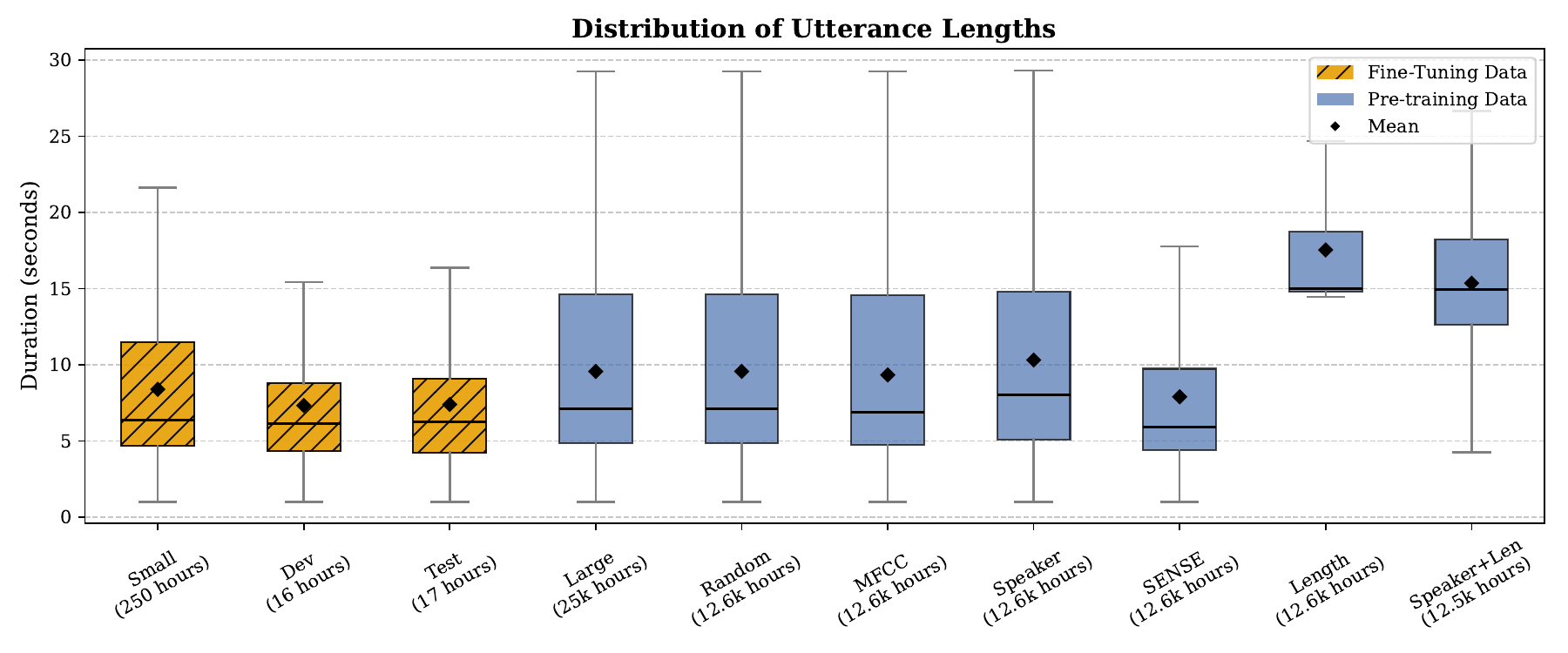}}
\caption{Box-and-whisker plots of utterance length distributions across different dataset splits. The three plots on the left correspond to the fine-tuning split, while the remaining plots show the various pre-training subsets. Notably, the best performing models were pre-trained on subsets consisting of longer utterances (far right two plots), whose distributions differ most from the fine-tuning data.}
\label{fig:dataset-dist}
\end{figure*}


The results are presented in Table~\ref{tab:results}.
Overall, diversity-based sampling methods did not yield significant improvements over the either baseline. The only exception is the speaker-based method on the large split, which achieved a word error rate of 17.97 on the test set—significantly better than the random baseline (18.54) but not the \emph{all} baseline (18.08).

In contrast, utterance length–based methods consistently outperformed both baselines. On the medium split, selecting the longest utterances reduced the word error rate to 19.02 on the test set, while combining speaker diversity with utterance length further improved performance to 18.97. On the large split, these same methods reached 17.77 and 17.42 word error rates, respectively, both significantly better than both baselines. Although the combined method performed best overall, the gains over length-only sampling were relatively small.

In terms of efficiency, training times for the medium split were similar across all methods. However, on the large split, length-based sampling substantially reduced pre-training time. Compared to the \emph{all} baseline, these methods trained about 24\% faster. This improvement stems from the dynamic batching strategy: with longer utterances, batches contain fewer examples, which lowers per-step computational cost.

In the Loquacious dataset, prior experiments were conducted with models of similar size but using a conformer encoder–decoder architecture trained fully supervised on the small split~\cite{parcollet25_interspeech}. While this setup is not directly comparable, it provides a useful reference: the 100M parameter supervised model trained on 250 hours of labeled data. Notably, all of our data selection methods outperform this baseline, despite relying on a less complex architecture. This demonstrates the benefits of self-supervised pre-training.

\section{Discussion and Conclusion}
From our results, we find that reducing the amount of pre-training data by sampling based on acoustic, speaker, or linguistic features does not yield improvements over a random baseline. In contrast, selecting subsets of longer utterances consistently leads to lower WER, despite these subsets being most out-of-distribution relative to the fine-tuning data. As shown in Figure~\ref{fig:dataset-dist}, the best-performing subsets contain utterances with median lengths around 15 seconds and mean lengths above 15 seconds, whereas other subsets fall between 5 and 10 seconds. 

While further work is needed to understand the underlying cause of this effect as well as the applicability to other SSL speech models and datasets, our findings suggest that pre-training benefits from longer utterances even when fine-tuning data is dominated by shorter segments. This is potentially due to these utterances having longer richer contexts for learning or more challenging learning signals.
In future work, we plan to address these limitations and extend this work to better understand what makes a good pre-training dataset.

In summary, we systematically studied simple unsupervised data selection methods for pre-training SSL speech models. We showed that enforcing diversity across acoustic, speaker, or linguistic features does not improve downstream ASR performance, whereas selecting the longest utterances achieves superior performance while using only half of the original dataset. Moreover, this approach reduces pre-training time by up to 24\% on large datasets. These results highlight utterance length as a key factor in constructing efficient pre-training datasets for SSL speech models.




\vfill\pagebreak
\newpage

\ninept
\bibliographystyle{IEEEbib}
\bibliography{strings,refs}

\end{document}